\documentclass[aps,manuscript,reprint,noshowpacs,noshowkeys,prd,balancelastpage,nofootinbib]{revtex4}
\usepackage[utf8]{inputenc}
\usepackage{color}
\usepackage{amsmath}
\usepackage{amssymb}
\usepackage{graphicx}
\usepackage[unicode=true,
 bookmarks=false,
 breaklinks=false,pdfborder={0 0 1},backref=section,colorlinks=true]
 {hyperref}
\hypersetup{
 linkcolor=blue,urlcolor=blue,citecolor=red}

\makeatletter

\providecommand{\tabularnewline}{\\}

\@ifundefined{textcolor}{}
{%
 \definecolor{BLACK}{gray}{0}
 \definecolor{WHITE}{gray}{1}
 \definecolor{RED}{rgb}{1,0,0}
 \definecolor{GREEN}{rgb}{0,1,0}
 \definecolor{BLUE}{rgb}{0,0,1}
 \definecolor{CYAN}{cmyk}{1,0,0,0}
 \definecolor{MAGENTA}{cmyk}{0,1,0,0}
 \definecolor{YELLOW}{cmyk}{0,0,1,0}
}

\usepackage{amsfonts}
\usepackage{mdframed}
\usepackage{footnote}
\usepackage{float}
\usepackage[font=footnotesize,it]{caption}
\usepackage{tikz}
\usepackage{tikz-3dplot}

\makeatother

\begin{document}
\title{Step Momentum Operator }
\author{M. Izadparast }
\email{masoumeh.izadparast@emu.edu.tr}

\author{S. Habib Mazharimousavi}
\email{habib.mazhari@emu.edu.tr}

\affiliation{Department of Physics, Faculty of Arts and Sciences, Eastern Mediterranean
University, Famagusta, North Cyprus via Mersin 10, Turkey}
\date{\today }
\begin{abstract}
In the present study, the concept of a quantum particle with step
momentum is introduced. The energy eigenvalues and eigenfunctions
of such particles are obtained in the context of the generalized momentum
operator, proposed recently in \cite{M.H1,M.H2}. While the number
of bound states with real energy for the particles with Hermitian
step momentum inside a square well is infinite, it is finite for a
particle with $\mathcal{PT}$ -symmetric momentum. 
\end{abstract}
\maketitle

\section{Introduction}

After several years of accreditation to Dirac's definition of Hermitian
quantum mechanics, the thought of non-Hermitian systems has been noted
by physicists \cite{Caliceti}. $\mathcal{PT}$-symmetric quantum
theory with real energy spectrum has been brought to attention by
Bender and Boettcher's proposal in \citep{Bender1} and prospered
during the last two decades. A $\mathcal{PT}$-symmetric operator
is a non-Hermitian operator which manifests the parity ($\mathcal{P}$)
and time-reversal ($\mathcal{T}$) symmetry. We note that $\mathcal{P}$
and $\mathcal{T}$ are defined as $\mathcal{P}x\mathcal{P}=-x$, and
$\mathcal{T}i\mathcal{T}=-i$. In a physical approach, a $\mathcal{PT}$-symmetric
Hamiltonian is recognized as a gain-loss system being in dynamical
equilibrium. Whereas, the time evolution of the corresponding wave
function is kept unitary and the continuity equation confirms the
conservation of the probability density \citep{MakingSense}. The
other substantial status to take care of is the reality of the eigenvalues.
A $\mathcal{PT}$-symmetric system gives real eigenvalues, under the
special circumstances, which $\mathcal{PT}$-symmetry of the system
is unbroken. However, when the $\mathcal{PT}$-symmetry is broken
then the eigenvalues are attained in complex conjugate pairs \citep{MakingSense,Dorey,BenderBook}.
Mathematically speaking, the domain of space coordinate is broadened
to the complex plane as we confront the complex potentials. Thus,
the conventional interpretation of inner product in Hilbert space,
$\mathcal{H}$, cannot conveniently describe comprehensive dynamical
models in non-Hermitian quantum mechanics. To tackle this issue, it
is a necessity to amend $\mathcal{H}$, mathematically, in the use
of physical systems. To do so, Mostafazadeh in \citep{Mostafazadeh}
has established the essential definitions in accordance with the pseudo-Hermitian
quantum physics which has been organized and reviewed in \citep{Mostafazadeh1}.
In \citep{Mostafazadeh,Mostafazadeh1}, it is indicated that a pseudo-Hermitian
Hamiltonian manifests the reality of the spectrum, yet, it is a necessary
and sufficient condition for a pseudo-Hermitian operator to be compatible
with an invertible and antilinear operator such as $\mathcal{PT}$-symmetric
Hamiltonian operator. Hence, $\mathcal{PT}$-symmetry does not guarantee
to have real eigenvalues. Accordingly, the pseudo-Hermitian Hamiltonian
is introduced as 
\begin{equation}
H^{\dagger}\Theta=\Theta H,\label{Pseudo-H}
\end{equation}
in which $\Theta$ is a linear pseudo-metric, positive-definite, Hermitian
and invertible operator defined by $\Theta=\Omega^{\dagger}\Omega$,
$\Theta:\mathcal{H}\rightarrow\mathcal{H}$. Upon \eqref{Pseudo-H}
and the definition of $\Theta$ one can modify the inner product 
\begin{equation}
\left(\phi,\psi\right)_{\mathcal{H}}=\left\langle \phi\mid\Theta\mid\psi\right\rangle .\label{inner-product}
\end{equation}
this states that $\Omega$ maps $\left|\psi\right\rangle $ from the
original Hilbert space $\mathcal{H}$ to the target Hilbert space
$\mathcal{L}$ resulting in the Hermitian Hamiltonian- with $\mid\psi\succ=\Omega\left|\psi\right\rangle $,
provided $\left|\psi\right\rangle \in\mathcal{H}$ and $\mid\psi\succ\in\mathcal{L}$.
Consequently, $H$ in $\mathcal{H}$ corresponds to the constructed
Hermitian Hamiltonian $\mathfrak{h}$ in $\mathcal{L}$ defined by
\citep{Mostafazadeh}
\begin{equation}
\mathfrak{h}=\Omega H\Omega^{-1}.\label{Hermitized}
\end{equation}
The importance of the Hilbert space modifications and non-Hermitian
quantum physics has inspired mathematicians and physicists to investigate
extensively in this realm to clarify the ambiguities. Correspondingly,
a remarkable triumph is collected in \citep{Znojilbook}, besides,
Znojil has elucidated the alternative approaches of different Hilbert
spaces in \citep{Znojil}.

A non-Hermitian Hamiltonian, i.e, $H\neq H^{\dagger}$, is constituted
mostly by the impose of a non-Hermitian potential energy, i.e., $V\left(x\right)\neq V^{*}\left(x\right)$.
While the contribution of kinetic energy consistently is unaltered
through the exclusivity of the canonical momentum operator, which
is Hermitian and by definition, i.e, $p=\frac{\hbar}{i}\frac{d}{dx}$,
is $\mathcal{PT}$-symmetric. Moreover, the idea of constructing Hamiltonian
using the generalized momentum operator opens the discussion in the
context of non-commutative relations associated with the extended
uncertainty principle (EUP). We recall that EUP is linked to the existence
of minimum momentum \citep{non-commutative1}. In the virtue of non-Hermitian
non-commutative quantum physics, the compatibility of the non-Hermitian
operators is discussed in \citep{Znojil-noncom}, while in \citep{non-commutative2}
it was shown that how the deformed canonical operators are connected
to the non-Hermitian quantum physics. In the recent investigation
\citep{non-comm-non-Her}, the non-Hermitian and non-commutative Hamiltonians
are transformed under a set of maps which are proposed based on the
implementation of the phase-space formalism.

Here in the present study, we construct a Hermitian step momentum
operator as well as its $\mathcal{PT}$-symmetric version in accordance
with the recipe presented in \citep{M.H1,M.H2}. Two examples, inspired
by the one-dimensional non-Hermitian square well (NSW) potential studied
by Znojil in \citep{Znojil1,Znojil2,Levi-Kovacs}, are presented.
The proposed momentum is founded upon inserting a discrete auxiliary
function into the formalism which clarifies the Hermiticity or $\mathcal{PT}$-symmetry
of the generalized momentum operator. These examples are considered
to be toy models for the more realistic step momentum operator. Let's
add that, classically step momentum implies a particle of constant
mass and given velocity, experiencing a change in its velocity in
an infinitesimally short time due to an impulsive external force. 

This paper is organized as follows. In Sec. II, we present the Hermitian
step momentum with the momentum eigenvalue problem. Then, we consider
the corresponding Schrödinger equation for a particle in an infinite
square well. In Sec. III, we carry out the identical processes for
the piece-wise $\mathcal{PT}$-symmetric version of the generalized
momentum. Finally in Sec. IV, the outcomes are summarized and compared
with the generic canonical momentum operator.

\section{hermitian step momentum operator}

In pursuit of investigating the generalized momentum operator, in
accordance with the formalism presented in \citep{M.H1}, we propose
a step Hermitian momentum given by
\begin{equation}
p=-i\hbar\left(1+\mu\left(x\right)\right)\partial_{x}-i\hbar\frac{d\mu\left(x\right)}{2dx}\label{momentum}
\end{equation}
in which the auxiliary function $\mu\left(x\right)$ is defined to
be
\begin{equation}
\mu\left(x\right)=\begin{cases}
\mu_{0} & x<0\\
-\mu_{0} & 0<x
\end{cases},\label{mureal}
\end{equation}
where $\mu_{0}\in\mathbb{R^{+}}$. Implementing the auxiliary function
into Eq. \eqref{momentum} leads to
\begin{equation}
p=\begin{cases}
-i\hbar\left(1+\mu_{0}\right)\partial_{x} & x<0\\
-i\hbar\left(1-\mu_{0}\right)\partial_{x} & 0<x
\end{cases}.\label{Hermom}
\end{equation}
Let's add that the momentum operator \eqref{Hermom} is Hermitian,
i.e, $p^{\dagger}=p$ and therefore one expects to attain real eigenvalues,
orthogonal eigenfunctions and positive norm. Next, we consider the
corresponding eigenvalue equation expressed by 
\begin{equation}
p\phi\left(x\right)=\mathfrak{p}\phi\left(x\right)\Rightarrow\begin{cases}
-i\hbar\left(1+\mu_{0}\right)\phi_{-}^{\prime}\left(x\right)=\mathfrak{p}\phi_{-}\left(x\right) & x<0\\
-i\hbar\left(1-\mu_{0}\right)\phi_{+}^{\prime}\left(x\right)=\mathfrak{p}\phi_{+}\left(x\right) & 0<x
\end{cases},\label{eigen-value-P}
\end{equation}
where $\mathfrak{p}$ is the eigenvalue of the eigenfunction $\phi_{\mathfrak{p}}\left(x\right)$
which is found to be
\begin{equation}
\phi_{\mathfrak{p}}\left(x\right)=N\begin{cases}
\exp\left(\frac{i\mathfrak{p}x}{\hbar\left(1+\mu_{0}\right)}\right) & x<0\\
1 & x=0\\
\exp\left(\frac{i\mathfrak{p}x}{\hbar\left(1-\mu_{0}\right)}\right) & 0<x
\end{cases},\label{eigenfunction}
\end{equation}
in which $N$ stands for the normalization constant. To have the solution
physically accepted, $\mathfrak{p}$ has to be real and due to no
additional constraint it is continuous. 

\subsection{Stationary Schrödinger Equation:}

\begin{figure}
\includegraphics[width=3.8in]{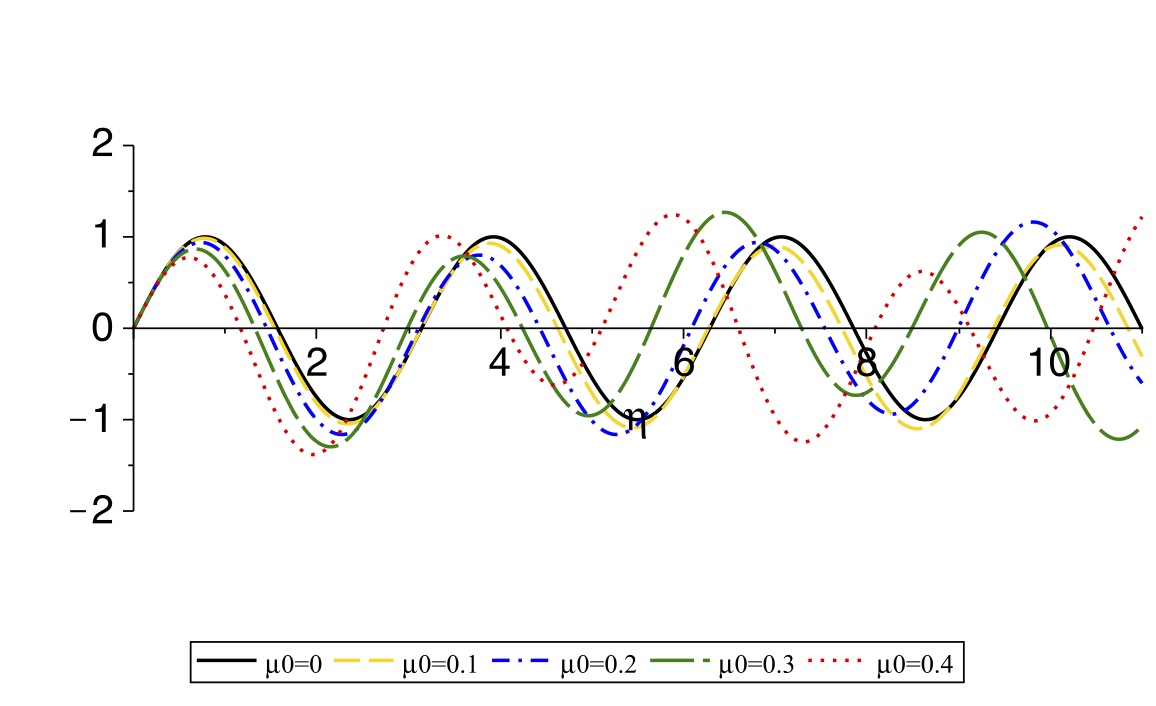}

\caption{Plot of Eq. \eqref{RealEnergy} with respect to $\eta$ for $\mu_{0}=0$
(solid, black), $\mu_{0}=0.1$ (dash, yellow), $\mu_{0}=0.2$ (dash-dot,
blue), $\mu_{0}=0.3$ (long-dash, green) and $\mu_{0}=0.4$ (dot,
red). The zero of each curve, corresponds to the $\eta_{n}=\lambda_{n}l$
in which $\lambda_{n}$ gives the allowed energy of the particle.
From this figure, we observe that, increasing the value of $\mu_{0}$,
decreases the energy of each corresponding state in comparison with
$\mu_{0}=0$.}

\label{FIG01}
\end{figure}

Upon applying Eq. \eqref{Hermom} into the Hamiltonian operator, one
finds the time independent Schrödinger equation in the following form
\begin{equation}
\begin{cases}
\frac{-\hbar^{2}}{2m}\left(1+\mu_{0}\right)^{2}\partial_{x}^{2}\psi\left(x\right)+V\left(x\right)\psi\left(x\right)=E\psi\left(x\right) & x<0\\
\frac{-\hbar^{2}}{2m}\left(1-\mu_{0}\right)^{2}\partial_{x}^{2}\psi\left(x\right)+V\left(x\right)\psi\left(x\right)=E\psi\left(x\right) & 0<x
\end{cases},\label{Ham-Hermitian}
\end{equation}
where the potential is assumed to be an infinite potential well defined
by
\begin{equation}
V(x)=\begin{cases}
0 & -l<x<l\\
\infty & l\leq|x|
\end{cases}.\label{Potential}
\end{equation}
It is appropriate to rewrite Eq. \eqref{Ham-Hermitian} in the form
\begin{equation}
\begin{cases}
\psi\left(x\right)^{\prime\prime}+\kappa^{2}\psi\left(x\right)=0 & -l<x<0\\
\psi\left(x\right)^{\prime\prime}+\bar{\kappa}^{2}\psi\left(x\right)=0 & 0<x<l
\end{cases},\label{Ham-simplified}
\end{equation}
where the double prime stands for the second derivative with respect
to $x$, $\kappa=\frac{\lambda}{\left(1+\mu_{0}\right)}$ , $\bar{\kappa}=\frac{\lambda}{\left(1-\mu_{0}\right)}$
and $\lambda^{2}=\frac{2mE}{\hbar^{2}}$. Eq. \eqref{Ham-simplified}
admits a solution expressed as
\begin{equation}
\psi\left(x\right)=\begin{cases}
Ae^{i\kappa x}+Be^{-i\kappa x} & -l<x<0\\
Ce^{i\bar{\kappa}x}+De^{-i\bar{\kappa}x} & 0<x<l
\end{cases},\label{wavef-Hermitian}
\end{equation}
in which $A,B,C$ and $D$ are integration constant. Next, we apply
the boundary and continuity conditions on the solution \eqref{wavef-Hermitian}
which lead to
\begin{equation}
\begin{cases}
\psi_{-}\left(0\right)=\psi_{+}\left(0\right),\\
\psi_{+}\left(-l\right)=0,\\
\psi_{-}\left(l\right)=0,\\
\partial_{x}\psi_{-}\left(0\right)=\partial_{x}\psi_{+}\left(0\right),
\end{cases}\Longrightarrow\begin{cases}
A+B-C-D=0\\
Ae^{-i\kappa l}+Be^{i\kappa l}=0,\\
Ce^{i\bar{\kappa}l}+De^{-i\bar{\kappa}l}=0\\
\kappa\left(A-B\right)-\bar{\kappa}\left(C-D\right)=0
\end{cases}.\label{Boundary-con}
\end{equation}
The latter is a homogenous system of four equations with four unknown
variables i.e.,$A,B,C$ and $D$. The system admits nontrivial solutions
if the determinant of coefficients vanishes. This, however, implies
\begin{equation}
\left|\begin{array}{cccc}
1 & 1 & -1 & -1\\
e^{-i\kappa l} & e^{i\kappa l} & 0 & 0\\
0 & 0 & e^{i\bar{\kappa}l} & e^{-i\bar{\kappa}l}\\
\kappa & -\kappa & -\bar{\kappa} & \bar{\kappa}
\end{array}\right|=0\label{determinant}
\end{equation}
 or explicitly 
\begin{equation}
\left(\bar{\kappa}-\kappa\right)\sin l\left(\kappa-\bar{\kappa}\right)+\left(\bar{\kappa}+\kappa\right)\sin l\left(\kappa+\bar{\kappa}\right)=0.\label{determinant-eq}
\end{equation}
 We note that $\kappa$ and $\bar{\kappa}$ contain the energy eigenvalue
in accordance with Eq. \eqref{Ham-simplified}. Hence, one can extract
the energy eigenvalues, numerically, using the latter equation. To
proceed solving \eqref{Boundary-con} provided $\kappa$ and $\bar{\kappa}$
satisfy the zero determinant condition \eqref{determinant-eq} we
introduce $b=\frac{B}{A},c=\frac{C}{A}$ and $d=\frac{D}{A}$ which
yields 
\begin{equation}
\begin{cases}
b-c-d=-1,\\
be^{i\kappa l}=-e^{-i\kappa l},\\
ce^{i\bar{\kappa}l}+de^{-i\bar{\kappa}l}=0,\\
-b\kappa-\bar{\kappa}c+\bar{\kappa}d=-\kappa
\end{cases}.\label{homogenous-eqs}
\end{equation}
The second equation clearly admits
\begin{equation}
b=-e^{-2i\kappa l}\label{eq1}
\end{equation}
which upon submitting into the other equations we obtain 
\begin{equation}
\begin{cases}
c=\frac{\kappa\cos\kappa l}{\bar{\kappa}\cos\bar{\kappa}l}e^{-i\left(\kappa+\bar{\kappa}\right)l}\\
d=-ce^{2i\bar{\kappa}l}
\end{cases}.\label{eq2}
\end{equation}
 The main equation which gives the legitimate discrete energy eigenvalues
i.e., \eqref{determinant-eq} and identifies the value of the wave
numbers $\kappa$ and $\bar{\kappa}$ gives also the constants of
integration with respect to $A$. The unknown constant $A$ will be
identified using the normalization condition. Let's introduce $\eta=\lambda l$
and recall the explicit definition of wave numbers which together
with equation \eqref{determinant-eq} imply
\begin{equation}
\left(1-\mu_{0}\right)\sin\left(\frac{\eta}{1-\mu_{0}}\right)\cos\left(\frac{\eta}{1+\mu_{0}}\right)+\left(1+\mu_{0}\right)\cos\left(\frac{\eta}{1-\mu_{0}}\right)\sin\left(\frac{\eta}{1+\mu_{0}}\right)=0.\label{RealEnergy}
\end{equation}
In Fig. \ref{FIG01}, we plot the left side of the latter equation
in terms of $\eta$ for various values of $\mu_{0}$. Also, in Tab.
\ref{Tab1} we present the energy eigenvalues of the first three bound
states of the system. We observe that, although, the momentum distribution
for negative and positive $x$ axis apparently cancel each other but
the energy eigenvalues decrease with increasing $\mu_{0}$. To see
the effect of the redistribution of the momentum operator on the probability
density we continue to find the explicit form of the wave functions.
To do so, we substitute the value of $B=bA,$ $C=cA$ and $D=dA$
in Eq. \eqref{wavef-Hermitian} and after some manipulation and redefinition
of the constant $A$, the eigenfunctions are obtained to be 

\begin{figure}
\includegraphics[scale=0.17]{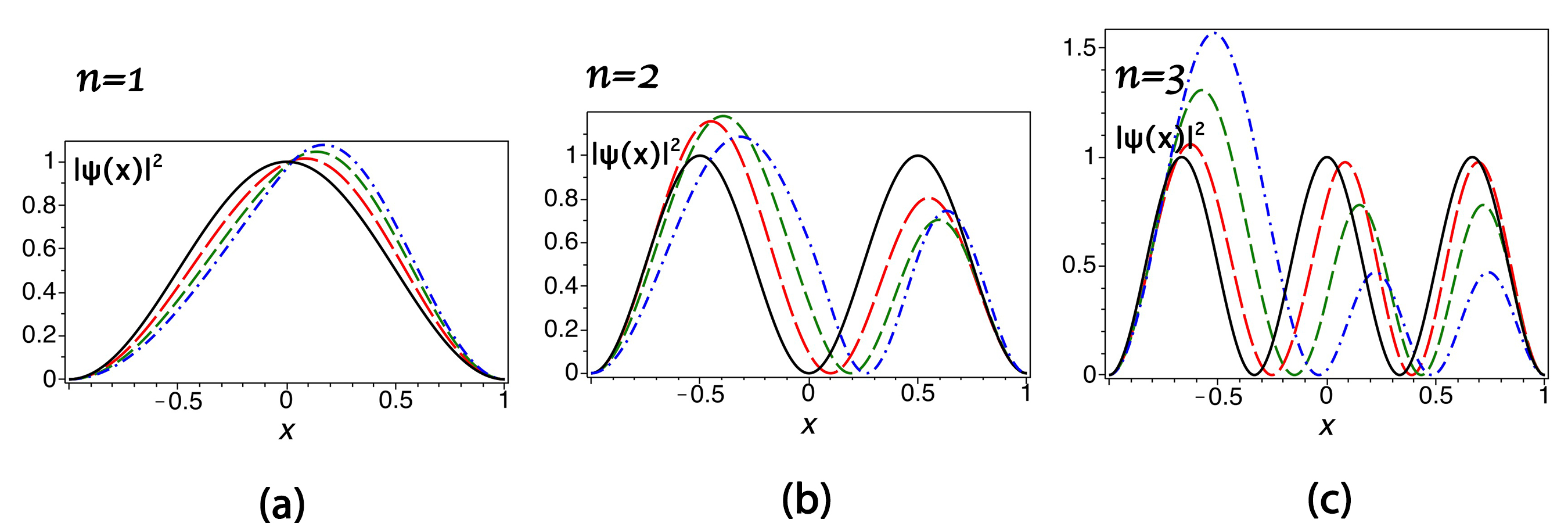}

\caption{Plots of the probability density of a quantum particle with Hermitian
step momentum in the ground, first excited and second excited states
in (a), (b), and (c) for $\mu_{0}=0$ (solid, black), $\mu_{0}=0.1$
(long-dash, red), $\mu_{0}=0.2$ (dash, green) and $\mu_{0}=0.3$
(dash-dot, blue).}

\label{FIG02}
\end{figure}

\begin{table}
\begin{tabular}{|c||c||c||c||c|}
\hline 
$\mu_{0}$ & \textbf{$0$} & \textbf{$0.1$} & \textbf{$0.2$} & \textbf{$0.3$}\tabularnewline
\hline 
\hline 
\textbf{$\frac{E_{1}}{E_{0}}$} & \textbf{$1$} & \textbf{$0.9621$} & \textbf{$0.8567$} & \textbf{$0.7112$}\tabularnewline
\hline 
\textbf{$\frac{E_{2}}{E_{0}}$} & \textbf{$4$} & \textbf{$3.9988$} & \textbf{$3.9211$} & \textbf{$3.6181$}\tabularnewline
\hline 
\textbf{$\frac{E_{3}}{E_{0}}$} & \textbf{$9$} & \textbf{$8.6796$} & \textbf{$7.9651$} & \textbf{$7.3006$}\tabularnewline
\hline 
\end{tabular}

\caption{The relative energy (i.e. $\frac{E_{n}}{E_{0}}$) for $n=1,2$ and
$3$ with $\mu_{0}=0.1,0.2$ and $0.3$. Note that $E_{0}$ is the
ground state energy of the system with $\mu_{0}=0$.}

\label{Tab1}
\end{table}

\begin{equation}
\psi_{n}\left(x\right)=N_{n}\begin{cases}
\sin\left(\bar{\kappa_{n}}l\right)\sin\left(\kappa_{n}\left(x+l\right)\right) & -l<x<0\\
-\sin\left(\kappa_{n}l\right)\sin\left(\bar{\kappa_{n}}\left(x-l\right)\right) & 0<x<l
\end{cases},\label{WaveF}
\end{equation}
in which $N_{n}$ is the normalization constant and the subindex $n=1,2,3,\ldots$
is the state number (the zero-number of \eqref{RealEnergy}) for each
given $\mu_{0}$. We would like also to add that the normalization
constant $N_{n}$ is function of state. In Fig. \ref{FIG02}, we plot
the probability density $\left|\psi_{n}\left(x\right)\right|^{2}$
for $n=1,2,3$ with different values of $\mu_{0}=0,0.1,0.2,0.3.$
It is observed that with increasing the value of $\mu_{0}$ the probability
distribution changes significantly such that the particle tends to
be in the negative $x$-axis. This is reasonable because by increasing
$\mu_{0}$ the momentum of the particle for $x<0$ is smaller than
for $x>0$ and the particle will have to spend more time in $x<0$. 

\section{$\mathcal{PT}$-Symmetric Step Momentum Operator}

\begin{figure}
\includegraphics[scale=0.3]{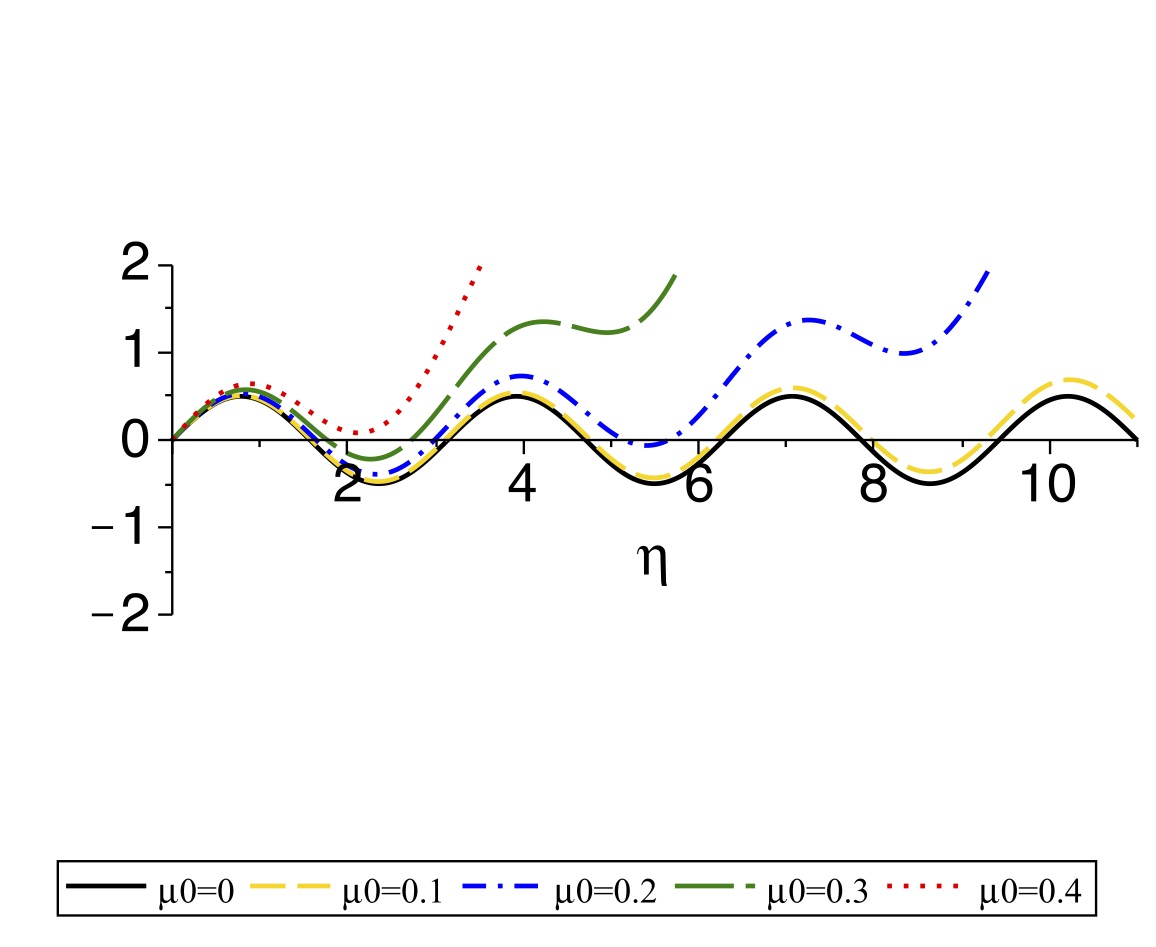}

\caption{Eq. \eqref{PT-energy} is plotted in terms of $\eta$, for $\mu_{0}=0$
(solid, black), $\mu_{0}=0.1$ (dash, yellow), $\mu_{0}=0.2$ (dash-dot,
blue), $\mu_{0}=0.3$ (long-dash, green), $\mu_{0}=0.4$ (dot, red). }

\label{FIG03}
\end{figure}

\begin{figure}
\includegraphics[scale=0.18]{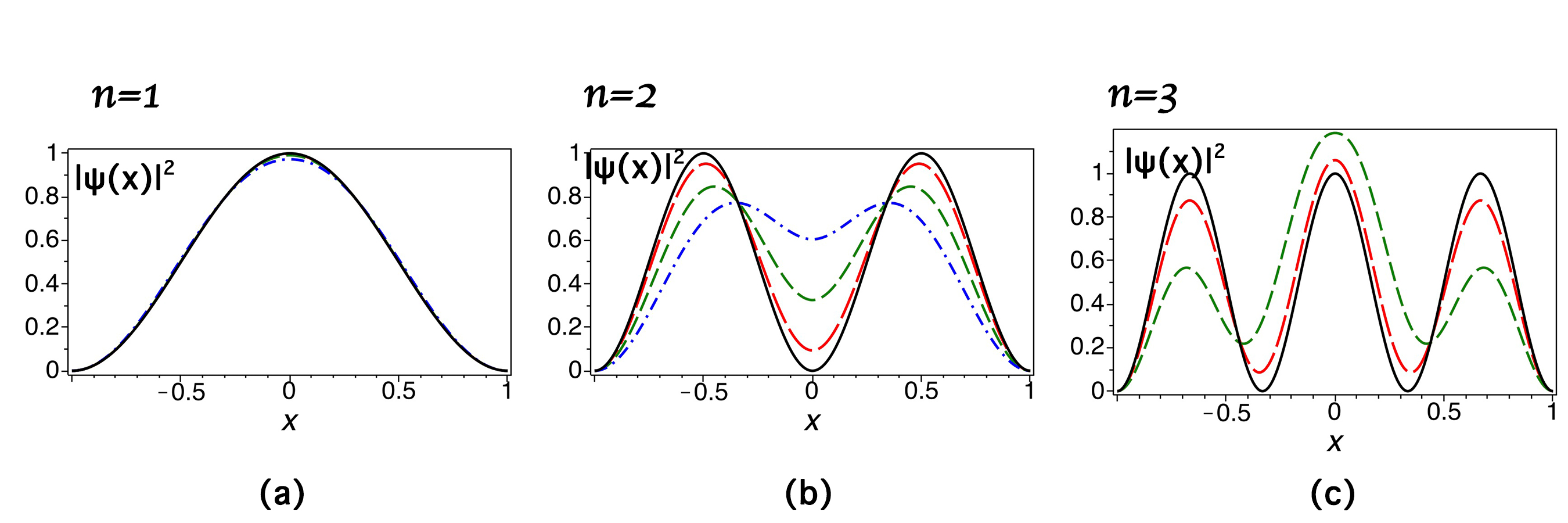}

\caption{Plots of $\left|\psi\left(x\right)\right|^{2}$- pseudo-probability
density- for a quantum particle with $\mathcal{PT}$-symmetric step
momentum inside an infinite square well representing the ground, first
excited and second excited states which are corresponded to $\mu_{0}=0$
(solid, black), $\mu_{0}=0.1$ (long-dash, red), $\mu_{0}=0.2$ (dash,
green) and $\mu_{0}=0.3$ (dash-dot, blue). }

\label{FIG04}
\end{figure}

In this section, we follow the same steps as of the previous section
and introduce a $\mathcal{PT}$-symmetric step momentum operator.
To do so, we set the auxiliary function $\mu\left(x\right)$ to be
in the form of a finite pure imaginary step function given by 

\begin{equation}
\mu(x)=\begin{cases}
i\mu_{0} & x<0\\
-i\mu_{0} & 0<x
\end{cases},\label{auxiliary-PT}
\end{equation}
in which $\mu_{0}$ is a positive real value. Inserting the latter
equation into Eq. \eqref{momentum} yields a $\mathcal{PT}$-symmetric
momentum operator expressed by
\begin{equation}
p=\begin{cases}
-i\hbar\left(1+i\mu_{0}\right)\partial_{x} & x<0\\
-i\hbar\left(1-i\mu_{0}\right)\partial_{x} & 0<x
\end{cases}.\label{momentum-PT}
\end{equation}
 Employing \eqref{momentum-PT}, one finds the time-independent Schrödinger
equation in the form of

\begin{equation}
\begin{cases}
\frac{-\hbar^{2}}{2m}\left(1+i\mu_{0}\right)^{2}\partial_{x}^{2}\psi\left(x\right)+V\left(x\right)\psi\left(x\right)=E\psi\left(x\right) & x<0\\
\frac{-\hbar^{2}}{2m}\left(1-i\mu_{0}\right)^{2}\partial_{x}^{2}\psi\left(x\right)+V\left(x\right)\psi\left(x\right)=E\psi\left(x\right) & 0<x
\end{cases},\label{Sch.PT}
\end{equation}
where $E\in\mathbb{R}$ is the energy eigenvalue of the eigenfunction
$\psi\left(x\right)$. Our choice of the potential $V\left(x\right)$
is an infinite square well given in Eq. \eqref{Potential}. Hence,
the Schrödinger equation turns to 
\begin{equation}
\begin{cases}
\psi\left(x\right)^{\prime\prime}+\kappa^{2}\psi\left(x\right)=0 & -l<x<0\\
\psi\left(x\right)^{\prime\prime}+\tilde{\kappa}^{2}\psi\left(x\right)=0 & 0<x<l
\end{cases},\label{Sch-PT-simple}
\end{equation}
in which $\kappa$ and $\tilde{\kappa}$ are defined to be $\kappa^{2}=\frac{2mE}{\hbar^{2}\left(1+i\mu_{0}\right)^{2}}=\left(s-it\right)^{2}$
and $\tilde{\kappa}^{2}=\frac{2mE}{\hbar^{2}\left(1-i\mu_{0}\right)^{2}}=\left(s+it\right)^{2}$.
Introducing a new real parameter 
\begin{equation}
\lambda=\sqrt{\frac{2mE}{\hbar^{2}}}\frac{1}{\left(1+\mu_{0}^{2}\right)},\label{eq:Lambda}
\end{equation}
 one obtains $s=\lambda$, $t=\lambda\mu_{0}$ and $\tilde{\kappa}$
and $\kappa$ are the complex conjugates of each other. From Eq. \eqref{Energy}
we may write
\begin{equation}
E=E_{0}\left(\frac{2\lambda l}{\pi}\right)^{2}\left(1+\mu_{0}^{2}\right)^{2},\label{Energy}
\end{equation}
with $E_{0}=\frac{\pi^{2}\hbar^{2}}{8ml^{2}}$. The solution to the
Schrodinger equation \eqref{Sch-PT-simple} is given by 
\begin{equation}
\psi\left(x\right)=\begin{cases}
Ae^{i\kappa x}+Be^{-i\kappa x} & -l<x<0\\
Ce^{i\tilde{\kappa}x}+De^{-i\tilde{\kappa}x} & 0<x<l
\end{cases},\label{WavePT}
\end{equation}
in which $A,B,C$ and $D$ are four integration constants. We note
that the boundary conditions for the new configuration, i.e., the
$\mathcal{PT}$-symmetric momentum operator other than Hermitian momentum
operator, are the same as Eq. \eqref{Boundary-con}. Therefore, the
similar proceedings in \eqref{determinant-eq} to \eqref{homogenous-eqs}
leads to a constraint equation in terms of $\eta=\lambda l$ expressed
by
\begin{equation}
\sin\left(2\eta\right)+\mu_{0}\sinh\left(2\eta\mu_{0}\right)=0.\label{PT-energy}
\end{equation}
We would like to add that the complex version of equation \eqref{determinant-eq}
admits two constraint equations due to the real and imaginary part
of the equation. However, the imaginary part is satisfied trivially
due to our assumption of real energy. Eq. \eqref{PT-energy} is a
transcendental equation, therefore, we plot \eqref{PT-energy} in
terms of $\eta$ for several values of $\mu_{0}$ in Fig. \ref{FIG03}.
The energy eigenvalues are obtained numerically upon applying the
zero's of \eqref{PT-energy} where the energy of the first three states
are expressed in Tab. \ref{Tab2} for $\mu_{0}=0.1,0.2,0.3$. It is
observed in Fig. \ref{FIG03} that the number of bound states are
finite such that for $\mu_{0}>0.377$ there is no bound state with
real energy. 
\begin{table}
\begin{tabular}{|c||c||c||c||c|}
\hline 
$\mu_{0}$ & \textbf{$0$} & \textbf{$0.1$} & \textbf{$0.2$} & \textbf{$0.3$}\tabularnewline
\hline 
\hline 
\textbf{$\frac{E_{1}}{E_{0}}$} & \textbf{$1$} & \textbf{$1.0422$} & \textbf{$1.1824$} & \textbf{$1.5035$}\tabularnewline
\hline 
\textbf{$\frac{E_{2}}{E_{0}}$} & \textbf{$4$} & \textbf{$3.9988$} & \textbf{$3.9205$} & \textbf{$3.5791$}\tabularnewline
\hline 
\textbf{$\frac{E_{3}}{E_{0}}$} & \textbf{$9$} & \textbf{$9.4073$} & \textbf{$11.6578$} & \textbf{$-$}\tabularnewline
\hline 
\end{tabular}

\caption{The relative energy (i.e. $\frac{E_{n}}{E_{0}}$) is presented for
$n=1,2$ and $3$ with $\mu_{0}=0.1,0.2$ and $0.3$ for a quantum
particle with $\mathcal{PT}$-symmetric step momentum. Note that $E_{0}$
is the ground state energy of the system with $\mu_{0}=0$. The number
of states with real energy is finite as for $\mu_{0}=0.3$, there
exist merely two bound states.}

\label{Tab2}
\end{table}
 Following the numerical value for every possible bound state one
can in principle find the wave function up to a normalization constant.
In terms of $\kappa_{n}$ and $\tilde{\kappa}_{n}$ the wave function
is given by 

\begin{equation}
\psi_{n}\left(x\right)=N_{n}\begin{cases}
\sin\left(\tilde{\kappa_{n}}l\right)\sin\left(\kappa_{n}\left(x+l\right)\right) & -l<x<0\\
\sin\left(\kappa_{n}l\right)\sin\left(\tilde{\kappa_{n}}\left(x-l\right)\right) & 0<x<l
\end{cases},\label{Wavefunction-PT}
\end{equation}
in which $N_{n}$ is the normalization constant for the state number
$n=1,2,3,\ldots$. We note that $\psi\left(x\right)\psi\left(x\right)^{*}$
is not a probability density when the Hamiltonian is not Hermitian.
We refer to \cite{Complex correspondence principle} where it is indicated
that three conditions $\Im\left(\psi\left(x\right)\mathcal{PT}\psi\left(x\right)\right)dx=0$,
$\Re\left(\psi\left(x\right)\mathcal{PT}\psi\left(x\right)\right)>0$
and $\int_{C}\left(\psi\left(x\right)\mathcal{PT}\psi\left(x\right)\right)dx=1$,
in which $x$ is the complex plane of the particles coordinate. Therefore,
instead of $\psi\left(x\right)\psi\left(x\right)^{*}$, $\psi\left(x\right)\mathcal{PT}\psi\left(x\right)$
represents the probability density in complex plane and $\mathbb{C}$
is the specific path where the above conditions are satisfied. Here,
it is worth to mention that the $\mathcal{PT}$-symmetric is not exact
in this problem. This can be seen clearly if we assume the energy
to be complex where we will be able to find the wave function which
satisfies all the conditions. In other words, for every specific $\mu_{0}$
there are finite number of bound state with real energy and infinite
number of bound states with complex energy. For the sake of comparison,
we plot $\left|\psi\left(x\right)\right|^{2}$ in Fig. \ref{FIG04}
for the first three bound states with different $\mu_{0}$. The effect
of $\mu_{0}$ can be seen as redistribution of $\left|\psi\left(x\right)\right|^{2}$-
We also note that $\int\left|\psi\left(x\right)\right|^{2}dx=1$.

\subsection{Remark}

Here, in this section we intend to find a relation between our outcomes
and the results which is acquired in Ref. \cite{Znojil1}. M. Znojil,
in Ref. \cite{Znojil1}, encounters with the ambiguity of continuation
of the $\mathcal{PT}$-symmetric square well potential. It is assumed
that the energy is real and defined as $E=t^{2}-s^{2}$ with $Z=2st$
to be the measure of non-Hermiticity. However, we rename the latter
energy as $E_{z}$ and the energy obtained in the present study $E_{\mu}$,
the subindices $Z$ and $\mu$ stand for the factor representing the
non-Hermiticity in both studies. Considering $\hbar^{2}=2m=1$, the
wave numbers are given by 
\begin{equation}
\begin{cases}
\kappa_{z}^{2}=E_{z}-iZ, & \kappa_{z}^{*2}=E_{z}+iZ\\
\kappa_{\mu}^{2}=\frac{E_{\mu}}{\left(1+i\mu_{0}\right)^{2}}, & \tilde{\kappa}_{\mu}^{2}=\frac{E_{\mu}}{\left(1-i\mu_{0}\right)^{2}}
\end{cases}.\label{waveNumbers}
\end{equation}
Having the identical formulation corresponding to the wave functions
in Eq. \eqref{wavef-Hermitian} and Eq. (6) in \cite{Znojil1}, we
suppose the wave numbers are equal, therefore, from the equality of
the latter equations one finds 
\begin{equation}
\begin{cases}
E_{z}-\frac{\left(1-\mu_{0}^{2}\right)}{\left(1+\mu_{0}^{2}\right)^{2}}E_{\mu}=0\\
Z-\frac{2\mu_{0}}{\left(1+\mu_{0}^{2}\right)^{2}}E_{\mu}=0
\end{cases}.\label{EqualWaveNumbers}
\end{equation}
Utilizing equations in \eqref{EqualWaveNumbers} with $Z,\mu_{0}>0$,
the relationship between $\mu_{0}$ and $Z$ admits
\begin{equation}
\mu_{0}=\frac{E_{z}}{Z}\left(-1+\sqrt{1+\frac{Z^{2}}{E_{z}^{2}}}\right).\label{mu}
\end{equation}
The physical approach of the latter comparison indicates that a particle
with the $\mathcal{PT}$-symmetric step momentum captured in a standard
infinite square well observes an effective potential of the form of
$\mathcal{PT}$-symmetric square well. This is due to the corresponding
variation in the non-Hermiticity factor in the momentum of the system.

\section{Conclusion}

Previously, we suggested a structure to form a generalized momentum
operator justifying the EUP which is an implication of the minimum
momentum. The paradigm was initially concerned with the Hermitian
operator \cite{M.H1}. Then, we modified the scheme to cover a wider
realm upon generating $\mathcal{PT}$-symmetric momentum operator
\cite{M.H2}. In the present study, we continued the investigation
around an exemplar developing a step momentum inspired by the infinite
non-Hermitian square well potential \cite{Znojil1}. First, we constructed
the so-called Hermitian step momentum operator. Its eigenvalue-problem
has been solved and the momentum eigenvalues and eigenfunctions have
been obtained. The corresponding Schrödinger equation of such particles
undergoing an infinite square well has been solved. We plot the probability
densities of some of the lower states to see the effect of the main
parameter $\mu_{0}$ in the distribution of the particle inside the
well. It is observed that the higher the step i. e., $2\mu_{0}$,
the more deviated distribution. This redistribution is in a way that
the particle tends to stay in the region with a smaller momentum operator
which is in agreement with our classical experience. In the second
part of the paper, we extended the concept of step momentum and introduced
the $\mathcal{PT}$-symmetric step momentum. The energy spectrum of
a particle with $\mathcal{PT}$-symmetric momentum inside a square
well has been calculated and we have shown that the number of bound
states with real energy is finite. Also, it indicates that the $\mathcal{PT}$-symmetry
of the system is broken. Besides, we plotted the pseudo-probability
density for a few lower states. The plots display the modification
of the probability density for $\mu_{0}\neq0$. We also found that
for $\mu_{0}>0.377$ there are no bound states with real energy.

\end{document}